# Bayesian Shape Invariant Model for Latent Growth Curve with Time-Invariant Covariates

Mohammad Alfrad Nobel Bhuiyan[1], Heidi Sucharew[2,3], Rhonda Szczesniak[2,3], Marepalli Rao[2,3], Jessica Woo[2,3], Jane Khoury[2,3], Md Monir Hossain[2,3]*

[1]Medpace Inc, 5375 Medpace Way, Cincinnati, OH, 45227

[2]Division of Biostatistics & Epidemiology, Cincinnati, Children's Hospital Medical Center, Cincinnati, OH, United States of America

[3]Department of Pediatrics, Cincinnati Children's Hospital, Medical Center, Cincinnati, OH, United States of America

**Abstract**

In the attention-deficit hyperactivity disorder (ADHD) study, children are prescribed different stimulant medications. The height measurements are recorded longitudinally along with the medication time. Differences among the patients are captured by the parameters suggested the Superimposition by Translation and Rotation (SITAR) model using three subject-specific parameters to estimate their deviation from the mean growth curve. In this paper, we generalize the SITAR model in a Bayesian way with time-invariant covariates. The time-invariant model allows us to predict latent growth factors. Since patients suffer from a common disease, they usually exhibit a similar pattern, and it is natural to build a nonlinear model that is shaped invariant. The model is semi-parametric, where the population time curve is modeled with a natural cubic spline. The original shape invariant growth curve model, motivated by epidemiological research on the evolution of pubertal heights over time, fits the underlying shape function for height over age and estimates subject-specific deviations from this curve in terms of size, tempo, and velocity using maximum likelihood. The usefulness of the model is illustrated in the attention deficit hyperactivity disorder (ADHD) study. Further, we demonstrated the effect of stimulant medications on pubertal growth by gender.

**Keywords:** Growth curve; Spline; Shape invariant; Longitudinal data

## Introduction

The analysis of childhood growth curves has played a vital role in estimating the growth trajectory of populations as well as identifying critical factors corresponding to various shapes of those trajectories, such as sex. Indeed, early origins of growth curve modeling utilized cross-sectional growth curves, in which the population data were used to derive the growth patterns for various age and gender groups. The two widely known references of using cross-sectional data are the CDC growth chart Kuczmarski [1] and WHO growth standards Organization [2], which are mainstays in clinical care. More recently, growth curve analyses based on longitudinal data have allowed more accurate identification of growth patterns, since the longitudinal data allows incorporating within- and between-subject effects simultaneously Willemsen et al. [3]. Earlier work fitted growth curves based on one of two parametric assumptions, namely logarithmic or exponential. Logarithmic curves assume a quick growth increase at the beginning, but the gains slowly disappear as time passes. Logarithmic growth curves broadly applied in bacterial growth Zwietering & Barry [4,5], biodegradation Schmidt et al. [6] fitness and strength training and learning ability. By contrast, the ex- potential curve assumes that growth is slower at the beginning with gains that are more rapid over time. Karlberg [7] explained growth curve modeling through exponential curves. Jenss [8] proposed a different model adding a linear term which was fitted by Berkey [9] and found a poor fit of the data based on the systematic variation of the residual Beath [10].

Moreover, Berkey [11] also analyzed the model proposed by Count Earl [12] and found an exponential model as a better model. Different parametric growth curve models were proposed over the last two decades, such as the linear model, reciprocal model, logistic model, Gompertz model, and the Weibull model. Milani & Wingerd [13,14] explored other approaches of growth curves and found a poor fit. Wishart [15] Explained growth data analysis using polynomial regression. One limitation of polynomial regression that it requires a higher degree of polynomial to provide an adequate fit with the resulting coefficients without having any significant interpretation. Alternative approaches to analyze growth curve data were proposed by Geva, Ong, Newell & Rao [16-19]. Nonparametric models using regression splines to model the underlying shape function have been shown to decrease bias, thereby improving the estimation of subject-specific effects Viele et al. [20]. Furthermore, regression splines, such as natural splines, have been shown to provide a better-localized fit to the mean response, compared to global polynomials Lambert et al. [21]. Equally important to finding an appropriate model to depict growth patterns is understanding the risk factors that contribute to adverse growth.

The growth model is essential for designing an intervention trial or increase public health awareness surrounding potential benefits and adverse effects of various environmental exposures and growth patterns. Further- more, having interpretable estimates for growth characteristics, such as peak height velocity, can ameliorate confounding in epidemiologic studies. Simpkin et al. [22]. Other examples, which have motivated statistical developments ranging from hierarchical linear models to multivariate analysis methods, include psychological change over time Hertzog [23], cardiovascular study Llabre et al. [24], associations between adolescent moderate-vigorous physical activity and depressive symptoms in young adulthood Brunet et al. [25], associations among the timing of sexual victimization and timing of drinking behavior Griffin et al. [26], examine longitudinal associations among cognition, function, and depression in Alzheimer's Disease patients Zahodne et al. [27], describing the change in personality trait Jackson et al. [28] In recent years, shape invariant modeling (SIM) has become an active area of research for nonparametric growth curve modeling, where a single function (or, curve) is transformed by scaling and shifting it to fit each subject usually through affine transformations. Lawton et al. [29], Who first proposed SIM called it self-modeling regression; in their approach, the function for the underlying shape illustrated for various parametric functions. Later, Beath [10] developed a model to explain longitudinal growth patterns and extended the SIM to include time-dependent covariates. Cole et al. [30] Extended the model by changing the sign of the velocity parameter and named it SITAR (Superimposition by Translation and Rotation). As a type of SIM, the regression spline expressed as a basis function consisting of a different set of knots; the resulting structure fitted as a nonlinear mixed-effects model and parameters typically estimated using maximum likelihood. This allows estimating the parameters for the between-subjects variation. Based on the underlying pattern of the data, various shape invariant models have proposed.

Bayesian growth curve modeling have also seen similar progress with many applications to real datasets as well as longitudinal growth datasets Barry [5], Arjas et al. [31-34]. One of the main advantages of a Bayesian approach is that it generates the uncertainty estimates (i.e., the estimate for the variance) for all unknown parameters naturally since each parameter explained by a probability distribution. Other advantages include the use of prior probability distributions to assimilate information from previous studies or expert's opinion and allows to control confounding; having posterior probabilities as easily interpretable alternatives to p-values; in hierarchical modeling, incorporating latent variables such as an individual's true disease status in the presence of a diagnostic error. Moreover, MCMC methodology facilitates the implementation of Bayesian analyses of complex data sets containing missing observations and multidimensional outcomes [31] Dunson et al. [32]. Due to this flexibility and better prediction of the exposure-outcome relation- ship, researchers are becoming more interested in Bayesian modeling. Notable work in Bayesian growth curve modeling includes the multivariate extension by Willemsen et al. [3] the original SIM model.

A brief outline of the paper is as follows. In Section 2, a short description of the original SIM model and the interpretation of various model parameters provided. Section 3 described the Bayesian implementation of the SIM model with and without subject-specific covariates, and also the DAG representations of these models. Part 4 illustrated the MCMC implementation, the specification for the prior distributions; the full conditional distribution and the posterior distribution for each model parameter; and how the assessment of model performance. The full derivation of posterior distributions given in the Appendix. Applications with real data provided in Section 5 and the Final Section includes the discussion and some proposals to the future extension.

### Shape Invariant Model (SIM)

Shape invariant model (SIM) is a non-parametric model where a single function is transformed by shifting and scaling to fit each subject. In SIM a regression spline is used as the function, with log transformation of the data. For this study we followed the notation from Beath [10], the SIM model can be expressed as:

$$y_{ij} = \gamma_{i2} + h\left(\frac{t_{ij} - \gamma_{i1}}{\exp(-\gamma_{i3})}\right) + s; \quad i_{ij} = 1,...N, and\, j = 1,...T$$

Here, the $y_{ij}$ is the growth measure of $i^{th}$ child at the $j^{th}$ time points which corresponds to age (in years) in our motivating example. Subject-specific coefficients $\gamma_i = (\gamma_{i1}, \gamma_{i2}, \gamma_{i3})$ enable each individual's growth trajectory to be aligned to a common growth curve, $h(\cdot)$, via transformations to the x- and y-axes. In this formulation, we will estimate $h(\cdot)$ using a spline function and let $s_{ij}$ be measurement error. The goal is to estimate the subject-specific vector $\gamma_i$ such that the corresponding individual growth curve from deviations from the average curve $h(\cdot)$. Following previously described work by Cole and others in equation (1), $\gamma_{i2}$ is termed as Size. $\gamma_{i2}$ can also be interpreted as subject-specific shift up or down in the spline curve along the response axis. $\gamma_{i2}$ is a random intercept term; when the response measure is height, $\gamma_{i2}$ is larger for taller children and smaller for shorter children. $\gamma_{i1}$ termed Tempo, which is a random time intercept and corresponds to differences in the timing of the growth spurt. This subject-specific left-right shift in the growth curves positive for late puberty and negative for early. The scaling factor within the spline function, $\gamma_{i3}$, is termed Velocity and corresponds to differences in the duration of the growth spurt between individuals. The Velocity parameter shrinks or stretches the time scale [30].

**Bayesian implementation of shape invariant model**

The above SIM model presented in equation (1) can be written with basis representation as follows.

$$y_{ij} = \gamma_{i2} + Z^T \beta_{i(K+2)} + s_{ij};\ i=1,\ldots N, \text{and}\ j=1,\ldots T_i = 1$$

Where,

$$Z_{ij} = B(\exp(\gamma_{i3})(t_{ij} - \gamma_{i1}))$$

$\gamma i = (\gamma_{i1}, \gamma_{i2}, \gamma_{i3}), \gamma_i \sim N_3(0, \Sigma_\gamma)$, and $s_{ij} \sim N(0, \sigma^2)$

$Z_{ij}$ is the basis of the natural cubic spline, evaluated at ($\exp(\gamma_{i3})$ $(t_{ij}\text{-}\gamma_{i1})$). Thus, $z_{ijk}$ is a vector of length κ+2, and β is the regression coefficient vector of same length. Here, κ is the number of inner knots and 2 is for the boundary knots. so natural cubic spline has κ+2 independent coefficients. Subject-specific $\gamma_i$ is assumed to have multivariate normal of order 3 with 0 mean vector and $\Sigma_\gamma$ variance-covariance matrix. We assume that $s_{ij}$ is independently normally distributed with mean 0 and variance $\sigma^2$, and also independent of $\gamma_i$. The other distributions for $s_{ij}$ such as Student's-t can also be considered depending on the type of growth data.

With subject-specific covariates: In many growth curves analyses, the needs for the inclusion of covariates for better predictability, as well as for better explanation of growth mechanisms are essential. For example, it may be of interest to know how the gender difference affects the growth patterns, or how the medication at early age for a specific disease condition affects the growth at later ages specifically by size, tempo, and velocity. The subject-specific covariates can be included in the model specified for $\gamma_i$ with a non-zero mean vector. If we assume (p-1) subject-specific covariates, the mean and the variance of $\gamma_i$ becomes,

$$\gamma_i \sim N_3(AX_i, \Sigma_\gamma)$$

where,

$$A = \begin{pmatrix} \alpha_{11} \alpha_{12} \ldots \alpha_{1p} \\ \alpha_{21} \alpha_{22} \ldots \alpha_{2p} \\ \alpha_{31} \alpha_{32} \ldots \alpha_{3p} \end{pmatrix}, X_i = (x_{i1}, x_{i2}, x_{ip})^T, \text{ and } \Sigma_\gamma = \begin{pmatrix} \sigma_{12} \sigma_{12} \sigma_{13} \\ \sigma_{21} \sigma_{22} \sigma_{23} \\ \sigma_{31} \sigma_{32} \sigma_{32} \end{pmatrix}$$

The first column of the regression coefficient matrix A is for the intercept, and the remaining columns are for each (p-1) covariate coefficients. Similarly, the first column of the design matrix X contains the value 1. Directed acyclic graph (DAG): DAG is a graphical representation of a hierarchical model that shows how the observed data and the unobserved parameters are conditionally dependent on each other (Figure 1). In the graph, the circle indicates the stochastic node (or, the unobserved parameters that need to be estimated), and the rectangle indicates observed data or the hyper-parameters where they were assigned to fixed values apriori (Figure 1A). When the Bayesian hierarchical model has a complex dependency structure, DAG helps in better visualization of the model as a whole, and the derivation of the posterior distribution for each stochastic node. (Figure 2) shows the DAG of the SIM model without covariates (a), and with subject-specific covariates (b) (Figure 2A).

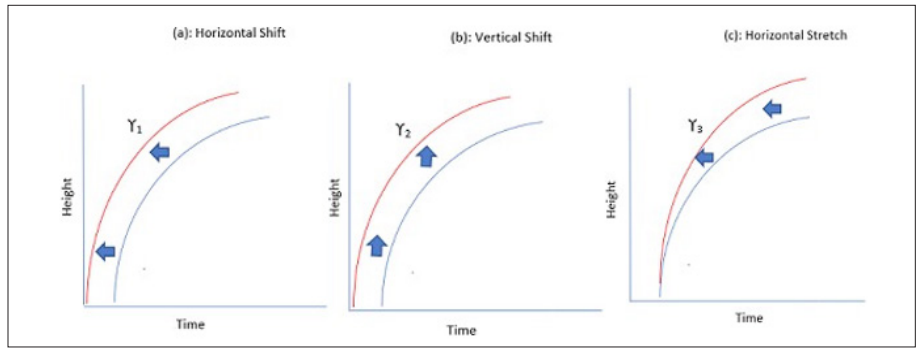

**Figure 1:** Schematic representation of shape invariant model with Horizontal shift (γ1), vertical shift(γ2) and stretch(γ3)

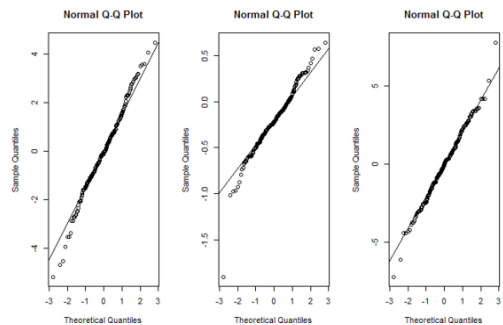

**Figure 1A:** Normality assumption of the subject-specific parameter for the subject-specific parameters (a) γ1 (b) γ2 and (c) γ3. The Q-Q plot indicates the normality assumption is satisfied.

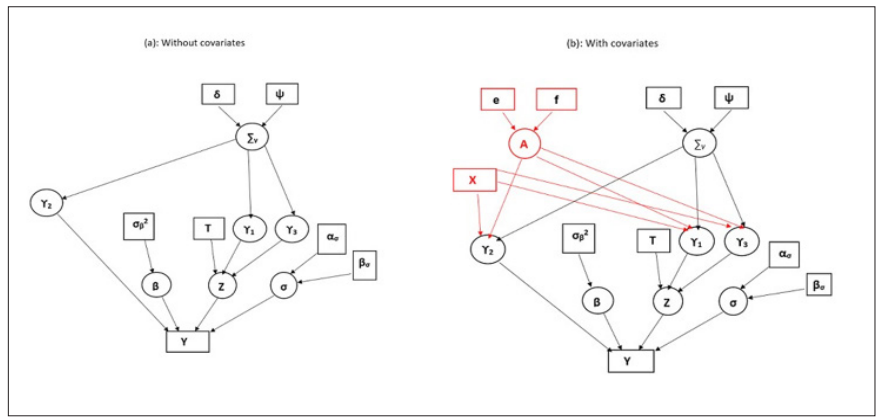

**Figure 2:** Graphical representation of the model without (a). and with subject-specific covariates (b). The subject specific covariates are shown in red.

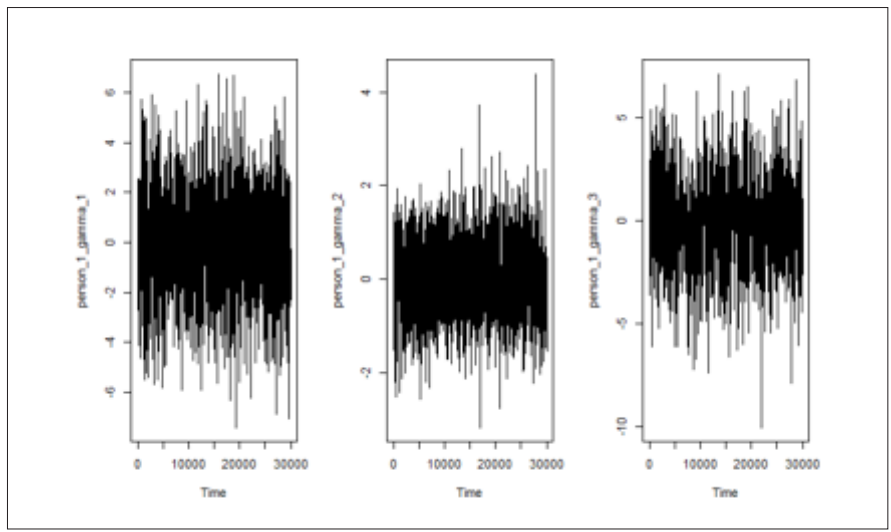

**Figure 2A:** Trace plot for the subject-specific parameters (a) γ1 (b) γ2 and (c) γ3. Trace plot indicates the convergence of the model parameters. We randomly selected a person and checked the trace plot, which indicates the model parameters converged.

**MCMC implementation**

**Prior distributions:** The prior distribution for all model parameters is assumed to be independent apriori and follow uninformative flat probability distribution in general. The residual variance was assumed to follow an inverse gamma distribution with fixed shape and scale parameters such that $\sigma^2 \sim IG(0.001, 0.001)$. Alternative prior distributions for the residual variance parameter can also be used following Gelman [35]. For covariate coefficients and basis coefficient, we have assumed $\alpha \sim N(0, 1000)$ and $\beta \sim N(0, 1000)$, respectively. The vector α was defined after stacking the A matrix. The variance-covariance matrix for $\gamma_i$ is assumed to follow an inverse Wishart distribution with 3 degree of freedom and 0.01 scale parameter, $\Sigma_\gamma \sim IW(3, 0.01)$.

**Full conditional distributions:** In Bayesian framework the joint posterior distribution is proportional to the product of the likelihood function and the prior distributions. Therefore, the full posterior distribution for the model in (2) with subject-specific covariates is as follows;

$$p(\alpha,\beta,\sigma^2,\gamma,\Sigma_\gamma \mid y) \alpha \prod_{i=1}^{N} \prod_{j=1}^{T_i} N(y_{ij} \mid \gamma_{i2} + z_{ij}(\gamma)^T \beta, \sigma^2) \prod_{i=1}^{N} N_3(\gamma_i \mid Ax_i, \Sigma_\gamma)$$

$$\prod_{i=1}^{3p} N(\alpha \mid 0, \sigma_\alpha I_{np}) \times \prod_{k=1}^{k+2} N(\beta_k \mid 0, \sigma_\beta^2) \times IG(\sigma^2 \mid \alpha_\sigma, \beta_\sigma) \times IW(\Sigma_\gamma \mid \delta, \psi) \quad (3)$$

We follow the block update procedures in MCMC iterations whenever the posterior distribution has a standard form. It improves the convergence and mixing as well, and also saves computational time. Following the Gibbs sampling procedure, the updates for each parameter are as shown below. For the sake of simplicity, covariate effects and subject-specific notation (i) are omitted for portions of the updates.

**Updating α:**

$$p(\alpha \mid \beta, \sigma^2, \gamma, \Sigma_\gamma, y) \alpha N(\overline{\Sigma}_\alpha [X^T \otimes \Sigma_\gamma^{-1} + 1_{np} \sigma_\alpha^2]^{-1}$$

Where, $\overline{\Sigma_\alpha} = [X^T X \otimes \Sigma_\gamma^{-1} + I_{np}\sigma_\alpha^2]$ and $\gamma_{vec}$ is a vector of stacked $\gamma^T$.

**Updating β:**

$$p(\beta | \alpha, \sigma^2, \gamma, \Sigma_\gamma, y) \alpha N(\sigma^{-2}\overline{\Sigma_\beta}Z^T(y-\gamma_2), \sigma^{-2}Z^TZ + \sigma_\beta^{-2}I_{k+2})$$

Where, $\overline{\Sigma_\beta} = [\sigma^{-2}Z^TZ + \sigma_\beta^{-2}I_{k+2}]$

**Updating σ²:**

$$p(\sigma^2 | \alpha, \beta, \gamma, \Sigma_\gamma, y) \alpha \text{ Inverse Gamma}\left(a + \frac{\Sigma T_i}{2}, b + \frac{\Sigma\{y_{ij} - \gamma_{i2} - B(\exp(\gamma_{i3}))(t_{ij} - \gamma_{i1}))\}^2}{2}\right)$$

**Updating Σγ (without covariate)**

$$p(\Sigma_\gamma | \alpha, \beta, \gamma, \sigma^2, y) \alpha \text{ Inverse Wishart}\left(\delta + n, (\psi^{-1} + \gamma^T\gamma)^{-1}\right)$$

where, δ is degrees of freedom and ψ is the scale matrix.

**Updating Σγ (with covariate)**

$$p(\Sigma_\gamma | \alpha, \beta, \gamma, \sigma^2, y) \alpha \text{ Inverse Wishart}\left(\delta + n, (\psi^{-1} + (\gamma - XA^T)^T(\gamma - XA))^{-1}\right)$$

where, δ is degrees of freedom and ψ is the scale matrix.

$$\gamma_i | y_i, x_i, \alpha, \beta, \sigma^2, \Sigma_\gamma \alpha N_3(\gamma_i | Ax_i, \Sigma_\gamma) \times \prod_j N(y_{ij} | \gamma_{i2} + B(\exp(\gamma_{i3}))(t_{ij} + \gamma_{i1}))\beta, \sigma^2))$$

Moreover, for the full conditional of the subject-specific parameter, $\gamma_i$ is given by,

For updating $\gamma_i$, we used a random walk Metropolis-Hastings (M-H) algorithm to generate posterior samples for subject-specific effect. The candidate samples are generated from multivariate Student's-t distribution with 5 degrees of freedom and mean at the current value. The variance parameter of multivariate Student's-t was appropriately tuned to ensure that the acceptance rate of candidate samples to posterior samples in M-H step is around 20-30%.

**Convergence, mixing, and identifiability**

Before summarizing the MCMC samples (or, posterior samples) as posterior mean, median, or highest posterior region, it is essential to check that the posterior samples for each parameter are converging, mixing well, and has less auto-correlation. The convergence of MCMC samples indicates how close we are to the actual posterior distribution and mixing suggests how well the parameter space explored. There are different ways of checking these, namely, trace plot, autocorrelation plot, QQ plot, Brooks plot, Gelman-Rubin test, etc. A simple exploration of the trace plot gives an insight into the characteristics of the MCMC samples. Trace plots are produced for each parameter and checked whether different starting values led to better mixing and convergence, saving the computational time as well.

The SIM model requires to specify the Spline function with a specific number of knots. We used both B-spline and the natural cubic spline functions for checking their relative performances. As for specifying the number of knots, we used five equally spaced knots for both spline function, three interior and two exteriors. The number of knots was determined based on a compromise between optimizing a fit criterion and the computational burden the prior distributions specified according to Section 4.1. The MCMC implementation followed a block update procedure with a mix of Gibbs and M-H algorithm. In an application, the inverse-Wishart distribution for the posterior distribution for Σγ redefined as a scaled inverse-Wishart distribution for correct estimation of correlation matrix as well as for quick convergence.

We ran multiple chains with a relatively more extended burn-in period. The model runs for 500,000 iterations with 90% burn-in samples. Then we have 50,000 samples for posterior inference. We further reduced the posterior samples to size 5,000 after thinning by parameter 10 for lowering the autocorrelation in posterior samples. All the results reported in this manuscript derived from these 5,000 posterior samples. Convergence was checked by examining the trace plot of the posterior samples for each parameter and by using the Gelman [36] test. We have also checked the autocorrelation from the autocorrelation plot, and it showed a much faster rate of decreasing towards zero with the increasing lag values. The variance- covariance matrix of the proposal density in the M-H algorithm was tuned accordingly so that the acceptance rate was approximately 20%.

Mixing well of posterior samples is a valuable property to ensure that there is no specific trend in posterior samples among the parameters, and they are independent as much as possible. We randomly selected three patients and their posterior samples for a horizontal shift ($\gamma_1$) and stretch ($\gamma_3$) parameters plotted in (Figure 3). The plot shows there is no specific trend in posterior samples for these two parameters, and they scattered around the center, indicating low correlation(Figure 3A).

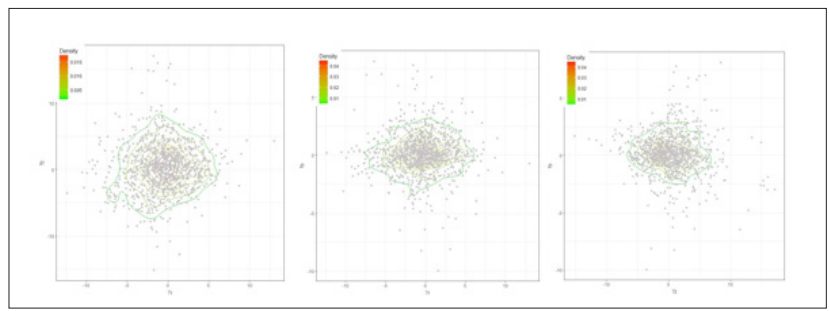

**Figure 3:** Contour plot of posterior samples for horizontal shift (γ1) and stretch (γ3) parameters for randomly selected three patients.

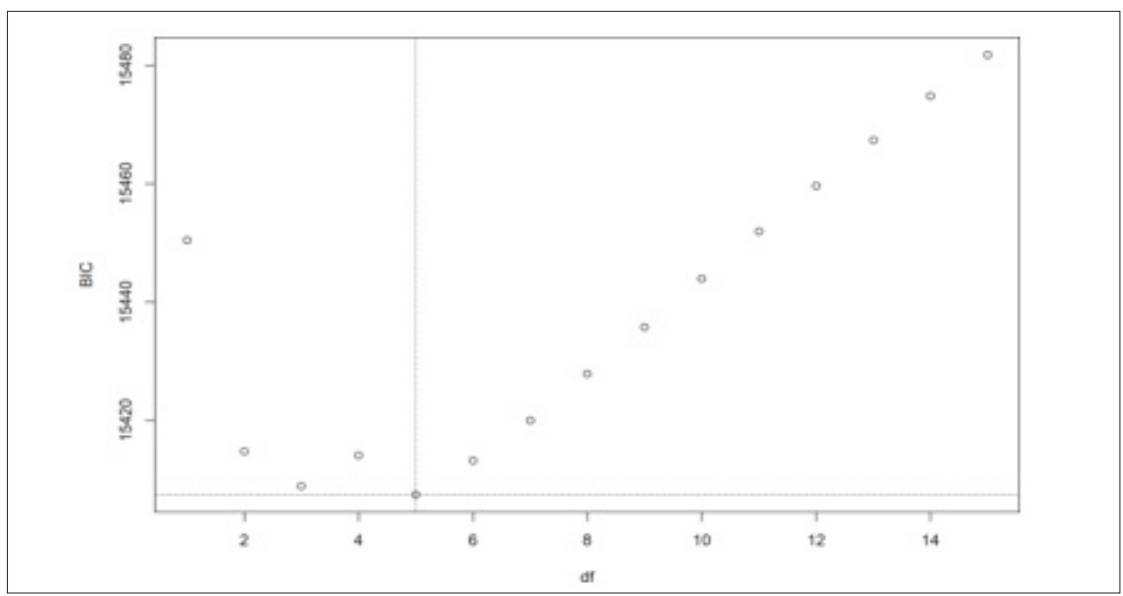

**Figure 3A:** We used the Bayesian information criterion to select the optimal number of knots. BIC plot for different number of knots.

**ADHD children data**

Longitudinal data from retrospective chart review for heights and weights for 197 ADHD (Attention-deficit Hyperactivity Disorder) children who visited community-based pediatric primary care practices in Cincinnati (Ohio, USA) collected. The children were in the age range: 5-16 years, and under stimulant medications to effectively reduce symptoms of ADHD. The objective of the original study was to evaluate how the age at start of stimulant medication may have impacted child growth trajectories. The original ADHD data had 6,134-time points recorded on 197 patients, among them only 3084 height measures were available for the analysis. The longitudinal study design is appeared to be unbalanced. The study enrolled 138 males (70%). For the 197 unique patients the age range was (5.02, 16.76) with the mean age 9.3 years, and the height range (76.2, 183.2) with the mean height 134.4cm. Each patient has prescribed a stimulant medication at a certain visit. The mean age at stimulant medication start was 7.9 years and ranged from 4.2 to 12.3 years. Descriptive statistics of the ADHD patient data is shown in (Table 1).

**Table 1:** Descriptive statistics of the ADHD patient data.

| Variable | Min | 1st.Q | Median | Mean | 3rd.Q | Max | SD |
|---|---|---|---|---|---|---|---|
| Age | 5 | 8 | 9.86 | 9.83 | 11.59 | 15.96 | 2.52 |
| Height | 76.2 | 122.6 | 134.6 | 134.4 | 147.3 | 183.2 | 18.86 |
| Wight | 9.072 | 21.32 | 28.12 | 31.71 | 38.33 | 114.3 | 31.51 |
| Start age of stimulant medication | 4.17 | 6.75 | 7.86 | 7.949 | 9.12 | 12.33 | 1.6 |

The height data were analyzed using Bayesian shape invariant growth curve model (equation 2) with and without covariates such that all individuals are assumed to have the same under- lying shape of the growth curve, subject to three simple transformations. This mean curve is estimated along with three subject-specific parameters termed size, tempo, and velocity that transform the mean curve to fit individual growth curves. The size parameter for each child shifts the fitted curve up/down, reflecting differences in size; the tempo parameter shifts it left/right, reflecting differences in puberty timing; and the velocity parameter stretches/shrinks the underlying age scale to make the curve shallower/steeper, reflecting differences in growth rate. The model fits the mean growth curve as a fixed effect natural cubic spline with specified degrees of freedom, and the parameters size, tempo, and velocity are estimated as subject- specific random effects. The models were found to fit better after log-transforming age, and the resulting coefficients can be multiplied by 100 and viewed as percentage differences. Figure 4 shows the height, shown (left) plotted as growth curves illustrating the sparsity of the ADHD patient data during puberty. We used different colors to separate the growth curves and the right plot indicates the plots after adjustment (Figure 4).

Looking at the individual curves, some children are consistently taller than average, and others always shorter (Figure 4A). Also, some start relatively early and become taller, while others do the opposite. Besides, all the children have a time when they grow appreciably faster than before or after called a pubertal growth spurt, and the timing of the spurt varies between 11 and 14 years. The shape invariant mean growth curve is estimated by taking the adjusted curves and fitting a natural cubic spline through them and shown in Table 2. The mean height velocity curve for male and female, calculated as the first derivative of the mean curve, is also shown in the Appendix. In general, a shift on the log scale translates to scaling on the original size. So, a left-right change on the log age scale corresponds to a shrinking-stretching on the age scale. Thus, for the log age model, the age scale can be viewed as elastic and fixed at zero, with the range shrunk for early puberty and stretched for later puberty. (Table 2) summarizes these effects, confirming that, for male children without medication were on average taller, later into adolescence, and less growing than earlier with medication children. The apparent exception is for girls' height, where the tempo effect (corresponding to age at peak weight velocity) was slightly first for medicated children. The velocity effect was dramatic, with growth in 1% faster for girls and 3% slower for males. It reflects a shape invariant model's attempt to reconcile materially different curve shapes in the two groups (Table 3). The mean growth curve for male and female and age at peak height is shown as a dashed vertical line in appendix A.4, and Subject-specific posterior mean estimates of γ with 95% credible interval is shown in appendix plot A.5.

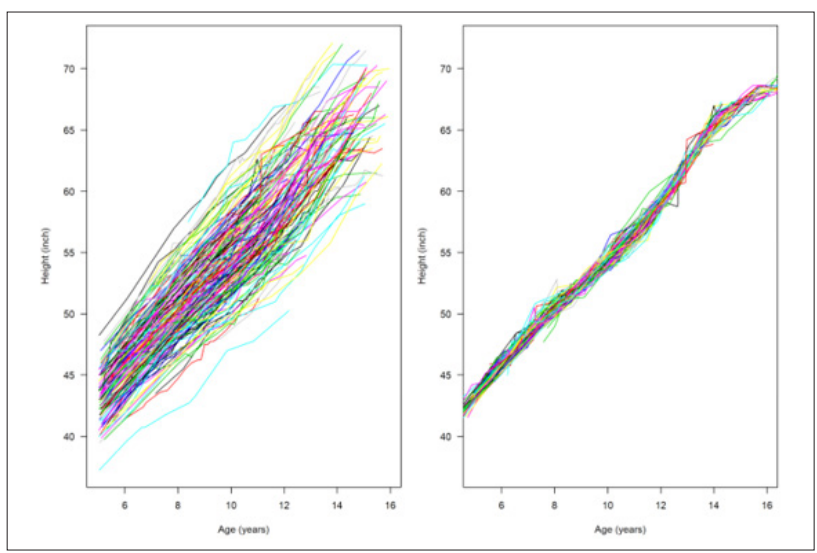

**Figure 4:** Age and Height curve data before and after adjustment.

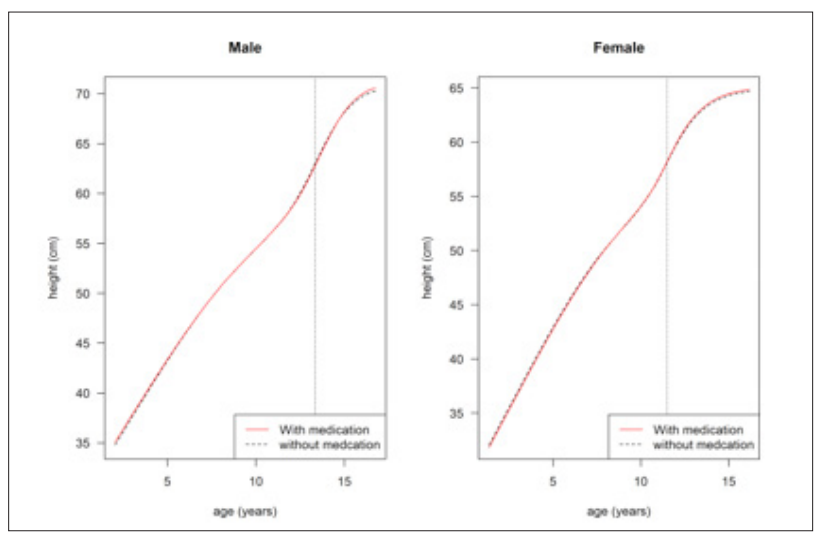

**Figure 4A:** we plotted the mean growth curve for Male and Female. The mean age at peak height is shown as a dashed vertical line. Mean growth curve plot indicates the mean growth difference between male and female after adjusting for age at first medication.

**Table 2:** Mean parameter estimates without and with covariate.

|  | Without Covariate | | With Covariate | |
|---|---|---|---|---|
|  | Male | Female | Male | Female |
| Size (inch) | 35.96 | 35 | 35.72 | 35.63 |
| Tempo | 2.2 | 0.48 | 2.33 | 0.47 |
| Velocity | -0.15 | 0.16 | -0.18 | 0.17 |

**Table 3:** Comparisons between growth parameters derived without and with age at first medication data, sexes combined.

|  | Without Covariate | With Covariate |
|---|---|---|
| Residual SD | 0.48 | 0.47 |
| SD of size | 3.55 | 3.24 |
| SD of tempo | 1.27 | 1.13 |
| SD of velocity | 0.11 | 0.1 |

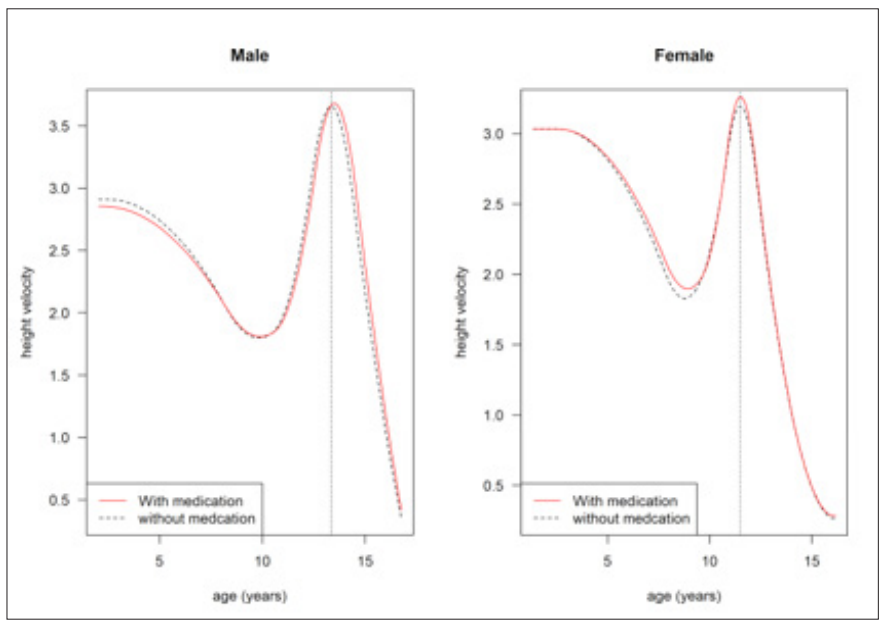

**Figure 5:** Mean growth curve with velocity for male(left) and female(right). Here the dotted vertical line indicates the velocity before adjusting for age at first medication variable and the red line indicates the mean growth curve after adjusting for the age at first medication variable.

The results of the models are affected by medication status. The mean age at a peak velocity of height is shown in Figure 5. Without medication, the age at peak velocity (APV) is 13.35 for males and 11.50 for females. With medication, the age at peak velocity (APV) cha and 11.51 for males and females, respectively. APV is marked by the vertical dotted line in the plots. The mean peak velocity of height without medication is 3.65 and 3.20cm/year for males and females, respectively, whereas the peak velocity changed to 3.68 and 3.27cm/year for males and females. Figure 5 shows age at peak velocity and peak velocity of the height data by sex for ADHD patients with and without medication status, illustrating the sparsity of the ADHD data during puberty.

**Table 4:** Comparisons between growth models with and without covariates.

| Model | Natural Cubic Spline | Basis Spline |
|---|---|---|
| Without Covariate | 19430.39 | 24718.74 |
| With Covariate | 12523.75 | 24693.43 |

Table 4 shows the fit of the models, in terms of the residual standard deviations (RSDs) and the SDs of the random effects, without and with age at first medication. In the without medication models, the RSD for height was more substantial than with the medication model for both sexes. Compared with without prescription, the tempo SD and the velocity SD are smaller in with the medication model. The results for the without and with medication models in Table 4 are broadly similar, showing that additional covariate effect on fit. It slightly reduced the RSD for height growth.

Correlating the random effects in the without medication models with those in the with medication models indicates that adding medication status increases the correlation between size and tempo for the male and female increase from 0.60 to 0.65 in boys and 0.31 to 0.33 in girls indicating improved estimates. The association between tempo and growth rate decrease 6% for male and 1% for female. Comparison of shape invariant tempo (the age they grow fast), Table 5 shows the correlations in the boys and girls between the measures of puberty timing and the shape invariant parameters from the with and without age at first medication models.

**Table 5:** Correlations of shape invariant parameters with puberty timing as obtained from the with and without Covariate.

|  | Without Covariate | | | | | | With Covariate | | | | | |
|---|---|---|---|---|---|---|---|---|---|---|---|---|
|  | Male | | | Female | | | Male | | | Female | | |
|  | Size | Tempo | Velocity | Size | Tempo | Velocity | Size | Tempo | Velocity | Size | Tempo | Velocity |
| Size | 1 | 0.6 | 0.32 | 1 | 0.31 | 0.3 | 1 | 0.65 | 0.3 | 1 | 0.33 | 0.34 |
| Tmpo | 0.6 | 1 | -0.37 | 0.31 | 1 | -0.57 | 0.65 | 1 | -0.43 | 0.33 | 1 | -0.58 |
| Velocity | 0.32 | -0.37 | 1 | 0.3 | -0.57 | 1 | 0.3 | -0.43 | 1 | 0.34 | -0.58 | 1 |

### Model selection

We implemented both the original SIM model and the Bayesian SIM model on the ADHD data. Two spline functions, B-spline and natural cubic spline used in the Bayesian SIM model. In all spline functions, five knots used. We checked the DIC values for the two spline functions for the Bayesian SIM model to test the model adequacy. Smaller the DIC is indicative of a parsimonious model. It appeared that the Bayesian SIM model with a natural cubic spline was performing better in both of these criteria (Table 4).

### Discussion

We have extended the SITAR model of Cole et al. [30] and incorporated the time-invariant covariate to analyze longitudinal growth curve data Stoel et al. [37]. The proposed model ensures that an individual's response curve adjusted to a standard shape curve, and that is a particular individual's response curve is some simple transformation of the usual shape curve. The model is semi-parametric when the population growth curve is modeled with a penalized regression spline. This model is useful in explaining subject-specific deviations in terms of size, tempo, and velocity from the underlying shape curve. This model estimates the standard error for estimates of subject-specific deviations for size, tempo, and velocity from the posterior samples for the subject-specific parameter ($\gamma$). Our model is flexible to assimilate information from previous studies or expert's opinions by specifying informative prior distributions. We implemented the model using different (e.g., Basis spline, natural cubic spline) spline functions and with an optimum number of knots on attention-deficit hyperactivity disorder data with and without time-invariant covariates to identify the effect of stimulant treatment on a growth spurt of children. We observed that our model with natural cubic spline function and time-invariant covariates has a better predictive ability. In our future works, we plan to extend the Bayesian

SIM model in finding clustering patterns in shape invariant parameters tempo ($\gamma_1$), size ($\gamma_2$), and velocity ($\gamma_3$). In many applications of shape invariant models in growth curve modeling, finding the group of subjects with similar growth patterns in terms of size, tempo, and velocity have real significance. Another extension can include but not limited to extending the model to free knot natural cubic spline, where we will utilize the Bayesian adaptive regression spline (BARS). It was shown earlier that BARS provides a parsimonious fit DiMatteo et al. [38].

### Acknowledgement

The project described was supported by the National Center for Advancing Translational Sci- ences of the National Institutes of Health, under Award Number 5UL1TR001425-04. The content is solely the responsibility of the authors and does not necessarily represent the official views of the NIH. Coauthor (RS) supported by grant K25 HL125954 from the National Heart, Lung, and Blood Institute of the National Institutes of Health.

### References

1. Robert JK, Cynthia LO, Shumei SG, Laurence MGS, Katherine MF (2002) 2000 CDC growth charts for the united states: methods and development. Vital Health Stat (246): 1-190.

2. World Health Organization (2006) WHO child growth standards: length/height for age, weight- for-age, weight-for-length, weight-for-height and body mass index-for-age, methods, and development. Geneva, Switzerland, pp. 1-312.

3. Sten PW, Paul HCE, Régine ST, Emmanuel L (2015) A multivariate bayesian model for embryonic growth. Statistics in Medicine 34(8): 1351-1365.

4. Zwietering MH, Jongenburger IL, Rombouts FM, Van't RK (1990) Modeling of the bacterial growth curve. Appl Environ Microbiol 56(6): 1875-1881.

5. Daniel B (1995) A Bayesian model for growth curve analysis. Biometrics 51(2): 639-655.

6. Steven KS, Stephen S, Martin A (1985) Models for the kinetics of biodegradation of organic compounds not supporting growth. Appl Environ Microbiol 50(2): 323-331.

7. Karlberg J (1987) On the modelling of human growth. Statistics in Medicine 6(2): 185-192.

8. Rachel MJ, Nancy B (1937) A mathematical method for studying the growth of a child. Human Biology 9(4): 556.

9. Catherine SB (1982) Comparison of two longitudinal growth models for preschool children. Biometrics 38(1): 221-234.

10. Ken JB (2007) Infant growth modelling using a shape invariant model with random effects. Stat Med 26(12): 2547-2564.

11. Catherine SB, Nan ML (1986) Nonlinear growth curve analysis: estimating the population parameters. Ann Hum Biol 13(2): 111-128.

12. Earl W (1943) Count growth patterns of the human physique: an approach to kinetic anthropometry: part I. Human Biology 15(1): 1-32.

13. Milani S, Bossi M, Marubini E (1989) Individual growth curves and longitudinal growth charts between 0 and 3 years. Acta Paediatrica 78(s350): 95-104.

14. Wingerd J (1970) The relation of growth from birth to 2 years to sex, parental size and other factors, using rao's method of the transformed time scale. Human biology 105-131.

15. Wishart J (1938) Growth-rate determinations in nutrition studies with the bacon pig, and their analysis. Biometrika 30(1/2): 16-28.

16. Diklah G, Lidush G, David S, Nancy LD (1993) A longitudinal analysis of the effect of prenatal alcohol exposure on growth. Alcohol Clin Exp Res 17(6): 1124-1129.

17. Ken KL, Michael AP, Pauline ME, Marion LA, David BD (2002) Size at birth and early childhood growth in relation to maternal smoking, parity and infant breast-feeding: longitudinal birth cohort study and analysis. Pediatr Res 52(6): 863-867.

18. Marie LN, Mario CB, Catherine P (2003) Height, weight, and growth in children born to mothers with hiv-1 infection in europe. Pediatrics 111(1): e52-60.

19. Radhakrishna CR (1958) Some statistical methods for comparison of growth curves. Biometrics 14(1): 1-17.

20. Kert V, Mark L, Robin LC (2006) Self-modeling structure of evoked postsynaptic potentials. Synapse 60(1): 32-44.

21. Paul CL, Keith RA, David RJ, Aidan WFH, Andrew S (2001) Analysis of ambulatory blood pressure monitor data using a hierarchical model incorporating restricted cubic splines and heterogeneous within-subject variances. Statistics in Medicine 20(24): 3789-3805.

22. Andrew JS, Adrian S, Mark SG, Jon H, Kate T (2017) Modelling height in adolescence: a comparison of methods for estimating the age at peak height velocity. Annals of Human Biology 44(8): 715-722.

23. Christopher H, John RN (2003) Assessing psychological change in adulthood: an overview of methodological issues. Psychol Aging 18(4): 639-657.

24. Maria ML, Susan S, Scott S, Patrice GS, Neil S (2004) Applying latent growth curve modeling to the investigation of individual differences in cardiovascular recovery from stress. Psychosom Med 66(1): 29-41.

25. Jennifer B, Catherine MS, Michael C, Tracie AB, Erin OL, et al. (2013) The association between past and current physical activity and depressive symptoms in young adults: a 10-year prospective study. Ann Epidemiol 23(1): 25-30.

26. Melissa JG, Jeffrey DW, Jennifer PR (2013) Recent sexual victimization and drinking behavior in newly matriculated college students: A latent growth analysis. Psychol Addict Behav 27(4): 966-973.

27. Laura BZ, Devangere PD, Yaakov S (2013) Coupled cognitive and functional change in Alzheimer's disease and the influence of depressive symptoms. J Alzheimers Dis 34(4): 851-860.

28. Joshua JJ, Patrick LH, Brennan RP, Brent WR, Elizabeth AL (2012) Can an old dog learn (and want to experience) new tricks? cognitive training increases openness to experience in older adults. Psychology and Aging 27(2): 286-296.

29. Lawton WH, Sylvestre EA, Maggio MS (1972) Self modeling nonlinear regression. Technometrics 14(3): 513-532.

30. Tim JC, Malcolm DC, Yoav BS (2010) Sitar? a useful instrument for growth curve analysis. Int J Epidemiol 39(6): 1558-1566.

31. Elja A, Liping L, Niko M (1997) Prediction of growth: a hierarchical bayesian approach. Biometrical Journal 39(6): 741-759.

32. Dunson DB (2001) Commentary: practical advantages of Bayesian analysis of epidemiologic data. American Journal of Epidemiology 153(12): 1222-1226.

33. José CP, Douglas MB (2000) Linear mixed-effects models: basic concepts and examples. Mixed-effects models in S and S-Plus, pp. 3-56.

34. Takao S, Kouji K, Takahiro S, TAO Qin (1991) A prediction of individual growth of height according to an empirical Bayesian approach. Annals of the Institute of Statistical Mathematics 43(4): 607-619.

35. Gelman A (2004) Prior distributions for variance parameters in hierarchical models. Technical report, EERI Research Paper Series.

36. Andrew G, Donald BR (1992) Inference from iterative simulation using multiple sequences. Statistical Science 7(4): 457-472.

37. Stoel RD, Wittenboer GVD, Joop H (2004) Including time-invariant covariates in the latent growth curve model. Structural Equation Modeling 11(2): 155-167.

38. Ilaria DM, Christopher RG, Robert EK (2001) Bayesian curve-fitting with free-knot splines. Biometrika 88(4): 1055-1071.